# 3D flow motions in the viscous sublayer

S. Santosh Kumar [1,2], Xinyi Huang[3], Xiang Yang[3] and Jiarong Hong[1,2*]

1. Department of Mechanical Engineering, University of Minnesota, Minneapolis, MN 55455
2. Saint Anthony Falls Laboratory, University of Minnesota, Minneapolis, MN 55414
3. Department of Mechanical Engineering, Penn State University, University Park, PA 16802

*Corresponding author: jhong@umn.edu

**Abstract**

We employ novel digital Fresnel reflection holography to capture the 3D flows within the viscous sublayer of a smooth-wall turbulent channel flow at $Re_\tau=400$. The measurements reveal unsteady and diverse flow patterns in the sublayer including nearly uniform high and low speed flows and strong small-scale (on the order of viscous wall units) spanwise meandering motions. The probability density functions (PDFs) of wall shear stresses show a clear discrepancy in high stress range with those from direct numerical simulation (DNS), which is attributed to the unresolved streamwise and spanwise motions by DNS. Moreover, the PDF of Lagrangian particle accelerations yields a stretched exponential shape like that in homogenous isotropic turbulence, indicating strong intermittency in the sublayer. We find a significant fraction of high accelerations is associated with the small-scale meandering motions. Our study helps explain the effect of sublayer-scale roughness on reducing drag and flow separation reported in the literature.

**Introduction**

The viscous sublayer is the region of turbulent flow closest to a surface, typically less than five wall units ($\delta_v$), where viscous effects are dominant. The fluid dynamics in the sublayer directly impacts wall shear stresses and therefore is relevant to the study of fundamental wall-bounded turbulence and its many applications. Using various visualization techniques based on passive tracers such as dust particles, dyes, and hydrogen bubbles, early investigations [1–3] revealed a qualitative picture of unsteady and organized fluid motions within the sublayer including spanwise "sinuous" motions, wall parallel streaks, etc. Subsequently, with the application of hot film anemometry (HFA), several studies investigated quantitatively the mean flow statistics within the sublayer [4–6] and highlighted important trends including the linear profiles of streamwise mean velocity, root-mean-square (rms) of the velocity fluctuation [4], and the increasing skewness and flatness of velocity distribution as the wall is approached [6]. To overcome the limitations associated with the intrusiveness of HFA, laser Doppler velocimetry (LDV) was implemented later to further examine the sublayer flow, which showed that the turbulence statistics within the sublayer are independent of the Reynolds number [7] and the intensity of velocity fluctuations at the wall asymptotes to a value smaller than the DNS prediction [8]. Despite these efforts, both HFA and LDV being single point measurements can only obtain spatially averaged flow quantities in a sample volume on the order of the viscous length scale (i.e., viscous wall unit $\delta_v$) or larger [7,9] and thus insufficient to resolve the flow structures within the sublayer.

To overcome these limitations, planar particle image velocimetry (PIV) and particle tracking velocimetry (PTV) were employed for sublayer studies [10–12]. Specifically, through single pixel correlations, Willert [10] resolved the velocity field and the instantaneous wall shear stresses and showed strong time symmetric correlation between the two. Li *et al.* [11] measured the streamwise velocity fluctuation in the sublayer and found it follows a logarithmic scaling with increasing Reynolds numbers. More recently, Willert *et al.* [12] captured rare events (less than 0.02% probability) of moving recirculation bubbles within the sublayer. However, the use of a light sheet with a thickness on the order of sublayer scale or larger in PIV and PTV can also lead to spatial averaging similar to those in HFA and LDV, limiting their effectiveness to quantify fine flow structures in the sublayer, particularly, those three-dimensional (3D) motions revealed qualitatively in the early investigations [1–3].

Acknowledgement of the fine flow structures in the viscous sublayer is vital towards understanding the physical mechanisms involved in the use of sublayer-scale roughness for flow control [13,14]. Specifically, surfaces with roughness embedded in the sublayer are generally considered to be hydrodynamically smooth [15]. Nevertheless, Sirovich and Karlsson [13] demonstrated that V-groove roughness ($5\delta_v$ tall, $200\delta_v$ wide) arranged with random shifts produces a reduction in skin friction, while using uniformly distributed cylindrical pillars (~$1\delta_v$ scale) with diverging tips, Evans *et al*. [14] observed a delay in flow separation in a channel with diverging cross section. To elucidate their mechanisms, a detailed quantification of the 3D flow structures within the sublayer at sufficient spatial and temporal resolutions is needed. However, the state-of-the-art 3D PIV and PTV techniques, including tomographic PIV/PTV [16–19] and holographic PTV [20,21], have only been implemented successfully for measurements in the logarithmic and buffer layers, owing to constraints in their optical setups and seeding approaches.

Therefore, in this study, we employ a novel in-house developed technique, i.e., digital Fresnel reflection holography (DFRH), for sublayer flow measurements. DFRH records holograms by interfering backscattered light from tracers with the reflection at the wall acting as the reference [22]. Such information is used to determine 3D tracer motions within the sublayer and the corresponding flow statistics. The results from our experiments are compared to a direct numerical simulation (DNS) performed at the same flow conditions. The experiments and DNS are performed on a smooth wall turbulent channel at $Re_\tau$~400, the complete details for which are included in the supplementary materials.

**Flow patterns within the viscous sublayer**

The tracer particle trajectories are captured in a sample volume of $20\delta_v$ (streamwise, *x*) × $20\delta_v$ (spanwise, *z*) × $12\delta_v$ (wall-normal, *y*) above the wall, with the closest trajectories measured at ~$0.1\delta_v$ to the wall. Figure 1 presents a compilation of trajectories captured in our experiments, based on which we can identify different flow patterns within the viscous sublayer. Specifically, a large fraction of the flow fields represented by the particle trajectories projected on to *x-y* plane shows a clear spatial variation of velocity within the sublayer with a sharp drop of streamwise velocity as the wall is approached (Figure 1b). In addition, the corresponding instantaneous velocity profiles do not exhibit a linear behavior with substantial variation along the spanwise direction (Supplementary Figure 3 and Supplementary Video 1), also indicated by the mixed high and low speed particle tracks at the same elevations in Figure 1b. In comparison to streamwise and spanwise motions, the wall normal motions are much weaker within our measurement volume (Figure 1b). The ensemble averaged inclination angle of particle trajectories (calculated as the arctangent of the wall normal displacement to streamwise displacement) is less than 3°, significantly smaller than the ones reported within the buffer and logarithmic layers (~11-30°) [17,23].

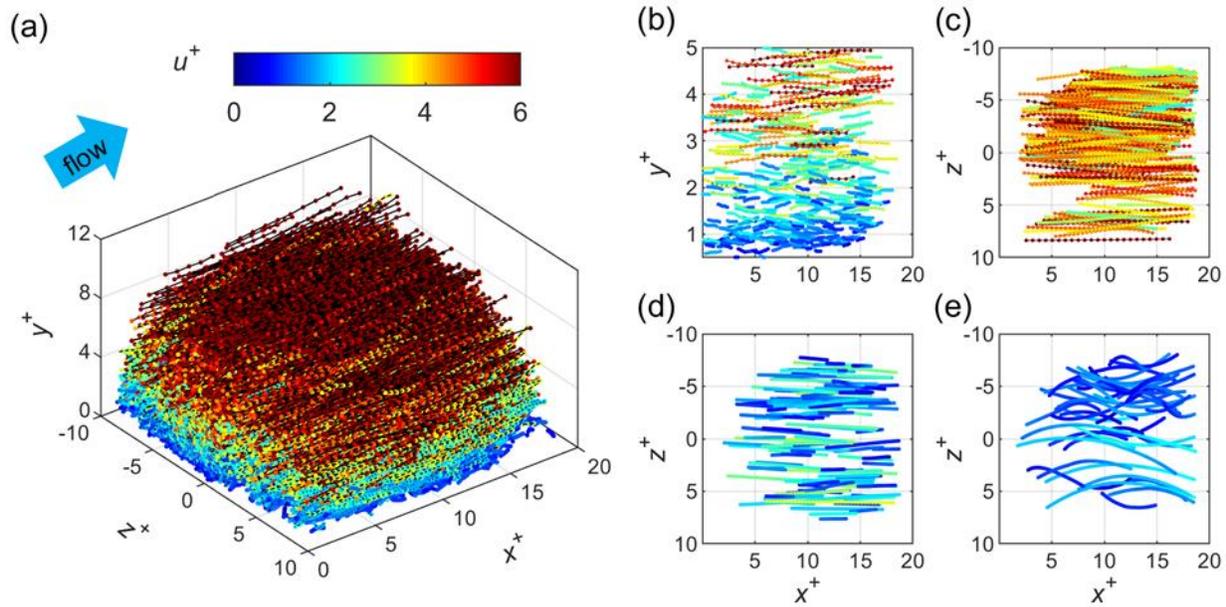

**Figure 1.** (a) A compilation of particle trajectories measured using DFRH in a sample volume of $20\delta_v$ (streamwise, $x$) × $20\delta_v$ (spanwise, $z$) × $12\delta_v$ (wall-normal, $y$) above the wall. In total about 250,000 trajectories obtained from nine datasets are shown here. Samples of particles trajectories below $y^+$~5 (where + indicates the quantity scaled by wall unit) illustrate different flow patterns in the viscous sublayer including (b) the flow with large velocity variation in the wall-normal direction ($x$-$y$ projection), nearly-uniform (c) high and (d) low speed flows ($x$-$z$ projection) as well as (e) spanwise meandering flows ($x$-$z$ projection). Note that the (b-d) are compilations of particle trajectories within a fixed time duration and (e) is obtained through a conditional sampling of trajectories with flow direction changing more than 30° from the start to the end of its trajectory as it moves across the sample volume, associated with the change in the spanwise velocity.

In addition, we observe the occurrence of nearly uniform high (Figure 1c) and low speed flows (Figure 1d) in the sublayer over 10% of the data. The temporal extent for these uniform high and low speed flows are estimated to be roughly within $\tau_I$ to $4\tau_I$, where $\tau_I$ is the integral time scale of the turbulent flow in our experiment, defined using hydraulic diameter of the channel and mean center line velocity. Based on their scales, these flow patterns are likely to be the footprints of commonly reported streamwise low and high speed streaks in the buffer layers, which yield temporal extents in a similar range [24,25].

Remarkably, our measurements also capture several instances of strong spanwise meandering motions (appearing in around 10% of the data), with spanwise scales on the order of ~2-3$\delta_v$ as illustrated in Figure 1e and Supplementary Video 2. Such near-wall meandering "sinuous" motions were reported by Fage and Townend [1], which used an ultramicroscope to examine the dust particles moving very near the wall in a water pipe. The location of the observation was at about 0.6% pipe radius away from the wall, likely to be within the sublayer based on their flow parameters. However, very few studies have since provided any further quantification from non-intrusive measurements of these motions within the sublayer. These meandering motions usually occur under low-speed flows deep within the sublayer, and some of these particle trajectories yield high curvatures, indicating strong accelerations (Supplementary Figure 4). The imprints of these motions on wall shear stresses and flow acceleration will be examined in detail in the following sections. It is worth noting that such meandering motions cannot be resolved by state-of-art DNS simulations, which have "standard" grid resolutions in the streamwise and spanwise directions of $12\delta_v$ and $5\delta_v$, respectively [26].

Overall, based on our observation, the flow patterns in the viscous sublayer are highly unsteady and spatially variable with dominant movements (both streamwise and spanwise) occurring on the wall-parallel planes. The burst and sweep events that involve strong movement in the streamwise wall-normal planes and are reported broadly in the existing measurements in the buffer layer [27] do not appear to be the dominant structures within the viscous sublayer. This trend is generally consistent with the asymptotic analysis of near-wall flow as the wall is approached [28]. As $y^+ \to 0$, the wall-normal velocity fluctuation varies as $(y^+)^2$ while the corresponding streamwise and spanwise values vary as $y^+$ resulting in a predominantly wall-parallel flow in the sublayer. Note that this asymptotic analysis holds for statistical quantities and does not preclude the presence of instantaneous motions with steep wall-normal components. Nevertheless, we do not observe such steep wall-normal motions in our data.

**Mean flow statistics within the viscous sublayer**

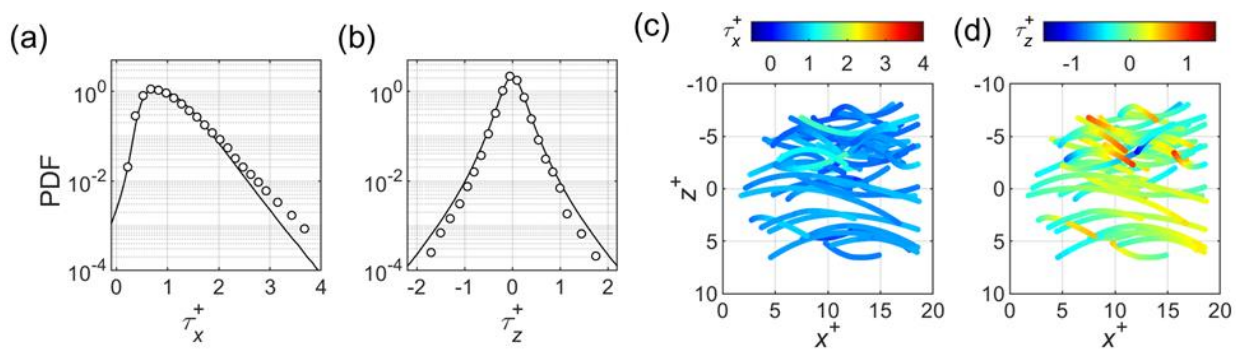

**Figure 2.** The probability density functions (PDFs) of the (a) streamwise and (b) spanwise wall shear stresses (○ representing experimental data) compared to the corresponding DNS predictions (solid lines). (c) Streamwise and (d) spanwise wall shear stresses associated with samples of the meandering motions that are currently unresolved by DNS with standard resolutions.

Based on the measured particle trajectories, the mean velocity profiles and the probability density functions (PDFs) of the streamwise and spanwise wall shear stresses are examined and compared with the DNS using standard grid resolutions. The mean streamwise velocity profile (Supplementary Figure 5a) shows a clear linear trend, matching both the DNS results as well as the hot-film [4–6] and laser doppler measurements [7,8] reported in literature. Correspondingly, the spanwise velocity (Supplementary Figure 5b) yields zero mean profile, consistent with the DNS predictions. This comparison provides a validation of DFRH techniques for turbulent flow measurements in the viscous sublayer. Moreover, neither of these mean profiles show any dependence on streamwise positions in our measurement regions, as shown in Supplementary Figure 6, suggesting the spatial variability observed in the flow patterns (Figure 1) is unlikely to be associated with the flow not being fully developed.

Following the same method of DNS, the instantaneous wall shear stress from our experiments (Figure 2) is calculated using the instantaneous velocity of the tracers deep within the viscous sublayer (i.e., $y^+$<1). As shown in Figure 2a, the PDF of streamwise wall shear stress components show a good agreement with DNS results in the lower range of positive stresses (i.e., $\tau_x^+ \lesssim 2$).

Remarkably, the experiment measurements deviate substantially from the DNS for higher stress events (i.e., $\tau_x^+ \gtrsim 2$), corresponding to two times (on average) difference in probability in this range. Similarly, for the PDF of spanwise wall shear stress (Fig. 2b), the matching between the experiments and DNS is also observed only for lower stress events (i.e., $|\tau_z^+| \lesssim 1$) and the deviation occurs for higher stresses (i.e., $|\tau_z^+| \gtrsim 1$) at close to the same level as that for $\tau_x^+$. The discrepancies between the experiments and DNS observed in the wall shear stress PDF are consistent with those in the profiles of streamwise and spanwise velocity fluctuations below $y^+ = 1$ (Supplementary Figure 5c and d), which show an increase of the streamwise and a decrease of the spanwise values compared to DNS. Despite the discrepancies present in the range of high stress events, the skewness values for the streamwise and spanwise wall shear stress measured in our experiments at 1.19, 0.07 are close (in terms of absolute values) to DNS estimates (0.99, -0.01, respectively) indicating a similarity in the overall shape of the distribution. However, the impact of these deviations on the kurtosis are more pronounced, with the experimental streamwise value larger than the DNS prediction (5.39 vs 4.88), and the corresponding spanwise estimate lower than the DNS value (7.36 vs 8.92). The smaller difference in the streamwise kurtosis between the experiment and DNS (compared with the spanwise kurtosis) may be attributed to the lack of reverse flow events observed in our experiment, which have been reported to occur at extremely low probabilities (~0.02%) in zero pressure gradient boundary layers [12].

It is worth noting that as with the mean velocity the wall shear stress PDFs show little dependence on streamwise position in our sample volume (Supplementary Figure 6). Therefore, we attribute these discrepancies to the motions unresolved by DNS, particularly, the small-scale spanwise meandering motions shown in Figure 1e. Figure 2c and d present the instantaneous streamwise and spanwise wall shear stresses associated with these meandering motions. As the figures show, these meandering motions yield low values of streamwise stresses while relatively high stresses along the spanwise direction. Note that the discrepancies observed between our experiments and DNS only suggest the lack of convergence of DNS (under standard streamwise and spanwise resolutions) to the real flows due to the presence of under-resolved motions. The PDFs of measured stresses being higher or lower than those from DNS do not imply the presence of higher or lower probabilities of large stress events in the experiments in comparison to the resolved events in DNS.

**Lagrangian particle acceleration within the viscous sublayer**

To understand the flow intermittency which is at the center of turbulence research [29] in the viscous sublayer, accelerations of tracer particles along their trajectories are examined within the sublayer. Despite the apparent quantitative difference, the PDF of Lagrangian particle acceleration shows some remarkable resemblance with the measurements from isotropic turbulence experiments [30,31]. Specifically, they both exhibit a stretched exponential shape with tails extending much higher than a Gaussian distribution of the same standard deviation, which suggests the existence of strong intermittency at small scales in both unbounded and bounded turbulent flows. In isotropic turbulence, such intermittency is known to be a result of uneven breakdown of a mother into two (or more) daughter eddies across the inertial range. Although the exact mechanism that leads to the intermittency in the viscous sublayer is still a research topic, the observation here lends support to the idea of momentum cascade, according to which the momentum flux is carried by wall-attached eddies of various sizes [32,33]. Similar to that in an isotropic turbulence, intermittency may naturally arise in the sublayer as a result of the uneven split of the momentum flux from a large-scale eddy to small scale eddies. It is worth noting that like the mean flow statistics, the observed acceleration statistics do not show appreciable dependence on streamwise positions within our sample volume (Supplementary Figure 7).

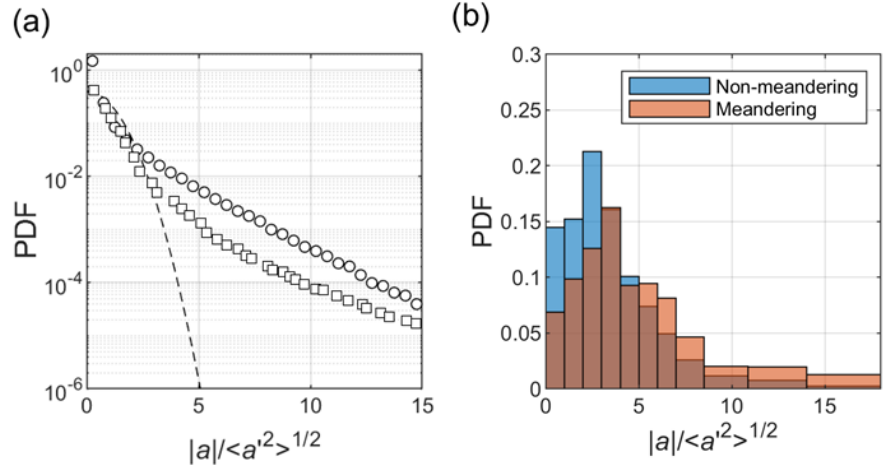

**Figure 3.** (a) The probability density function (PDF) of the magnitude of tracer particle acceleration (○) compared to a Gaussian distribution (dashed line) of the same standard deviation and Lagrangian acceleration measurements in isotropic turbulence (□) reported in Mordant *et al.* [31]. (b) Comparison of acceleration PDF for meandering vs non-meandering motions, where meandering motions are extracted from the data based on the same conditional sampling used in Figure 1e.

We attribute the high accelerations observed in Figure 3a to the presence of strong spanwise meandering motions within the sublayer. To illustrate this point, the acceleration PDFs corresponding to non-meandering and meandering motions are compared in Figure 3b. Note that the meandering motions are extracted using the same metric implemented for Figure 1e and the streamwise velocity for non-meandering motions is sampled to be within the same range as the meandering ones (i.e., $u^+ \lesssim 1.5$) for an accurate comparison. As the figure shows, the probability of high accelerations for the meandering motions are approximately twice as high relative to their non-meandering counterparts.

**Influence of sublayer roughness on macroscopic flow behaviors**

The flow structures in the viscous sublayer observed in our smooth-wall experiments can shed light on the impact of sublayer-scale roughness on macroscopic flow behaviors. Specifically, the sublayer-scale meandering flows (on the order of ~2-3$\delta_v$ observed in our experiments) accompanied with strong spanwise acceleration, indicates a significant contribution of spanwise motions to the energy dissipation through the strain rate tensor ($S_{ij}$). Such spanwise motions are potential sources of instabilities in the flow, which when acted upon by an external stimulus (e.g., adverse pressure gradient) can trigger flow separation, forming a recirculation bubble [3]. Therefore, sublayer-scale roughness can potentially create barriers to reduce the extent of spanwise motions and align flow from spanwise to the streamwise direction (Figure 4a). Such alignment would lead to an increase in the net streamwise flow velocity near the wall, suppressing the flow separation as shown by Evans *et al.* [14] in a diverging channel section (under adverse pressure gradient) coated with sublayer-scale pillars. Similarly, the presence of roughness in the sublayer may also influence the interaction between sublayer flow and overlaying large-scale turbulence structures, modulating their behaviors, and causing changes in the energy dissipation rate and thus drag on surfaces. In particular, as illustrated in Figure 4b and 4c, the near-wall large scale ($\sim 100\delta_v$) high and low speed streaks may interact with flow in the sublayer and leave signatures as uniformly high and low speed motions in our measurement regions (Figure 1b and c). Correspondingly, sublayer-height roughness with spacing and alignment matching these large-scale streaks can potentially stabilize this interaction and facilitate the momentum exchange between sublayer flow and overlaying turbulence, resulting in an increase in drag (Figure 4b). In contrast, the random arrangement of such roughness may disrupt sublayer signatures of these large-scale structures, destabilizing them and reducing the friction drag accordingly. This hypothesis can explain the behavior of V-groove roughness of $\sim 200\delta_v$ width and $\sim 5\delta_v$ height observed in Sirovich and Karlsson [13], which increases drag when the roughness is aligned with the mean flow but leads to drag reduction when randomly arranged. Nevertheless, further validation of our hypotheses requires direct measurements of flow around these sublayer roughness structures, which we plan to pursue using our DFRH technique in the future.

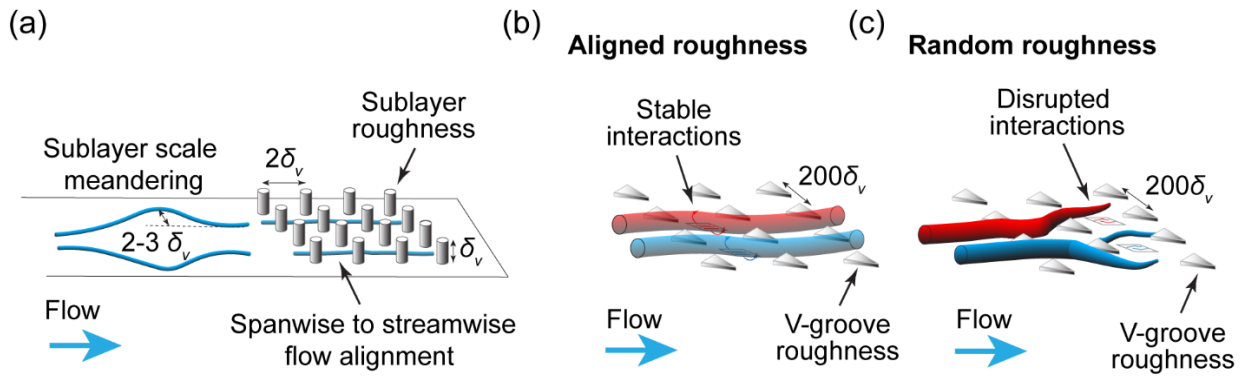

**Figure 4.** Schematics showing (a) the possible influence of roughness on sublayer-scale meandering flows in the viscous sublayer through flow alignment and the effects of (b) aligned and (c) randomly arranged V-groove roughness in the viscous sublayer on the potential interaction of sublayer flow with large-scale streaks above.

In the end, based on our sublayer measurement on a smooth wall channel, we would like to comment on the potential limitations of some near-wall measurement techniques, specifically, local seeding employed commonly for near-wall particle tracking velocimetry (PTV) [18,20,21]. The implementation of such seeding introduces a wall-normal component of velocity associated with the injection of tracer particles. In the literature, the injection speed is limited to a small fraction of centerline velocity in the test section to minimize the influence of local seeding on the near wall flow structures [34]. However, considering the small wall-normal velocity observed in our viscous sublayer measurements, the influence of such method needs to be considered carefully when studying the interaction between the sublayer flow and overlaying turbulence. As another example, shear stress sensors using micropillars, introduced by Brucker *et al*. [35] and extended for dynamic shear stress in Grosse *et al*. [36,37], are based on the assumption that the viscous sublayer flow is steady and laminar. The highly unsteady and spatially varying flow patterns observed in our sublayer measurements may cast some doubts on the accuracy of instantaneous shear stress measurements using such sensors in a turbulent near-wall flow. Specifically, their spanwise wall shear stress distribution has a kurtosis of 5.46 [37], substantially smaller than both our experimental measurements and the DNS prediction.

## Acknowledgments

Funding for this research was provided by The Office of Naval Research (Grant no. N000141612755). A large part of the computational work was performed using resources at the Minnesota Supercomputing Institute (MSI) at the University of Minnesota (http://www.msi.umn.edu) and the ACI at Penn State which the authors would also like to thank and acknowledge.

# 3D flow motions in the viscous sublayer: supplementary materials

S. Santosh Kumar [1,2], Xinyi Huang [3], Xiang Yang [3] and Jiarong Hong [1,2*]

1. Department of Mechanical Engineering, University of Minnesota, Minneapolis, MN 55455
2. Saint Anthony Falls Laboratory, University of Minnesota, Minneapolis, MN 55414
3. Department of Mechanical Engineering, Penn State University, University Park, PA 16802

*Corresponding author: jhong@umn.edu

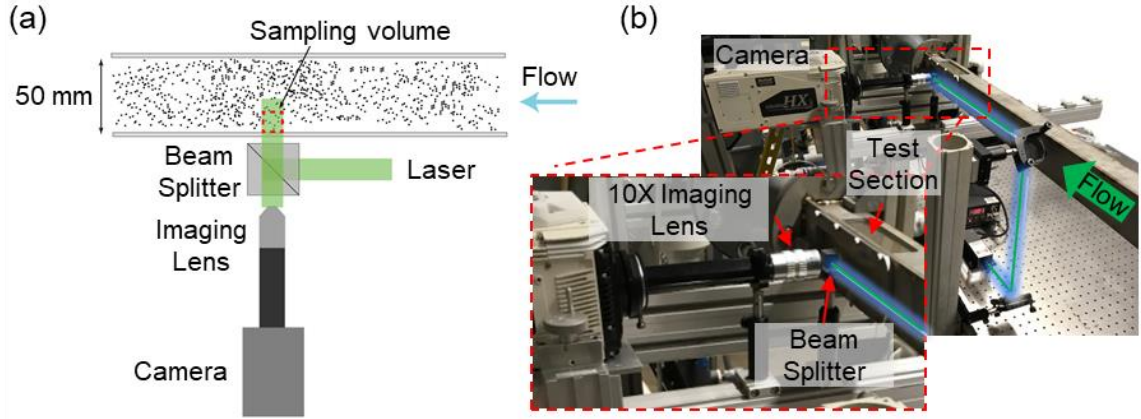

Supplementary Figure 1. (a) Schematic and (b) photo of the experimental setup of smooth-wall viscous sublayer flow measurements using Digital Fresnel Reflection Holography (DFRH).

The viscous sublayer measurements are conducted in the smooth-wall channel of a recirculating water tunnel using digital Fresnel reflection holography (DFRH) (Supplementary Figure 1a). The facility consists of a 1200 mm long, 50 mm × 50 mm square test section. The sample volume of the measurements is located at about 1100 mm downstream from the entrance of the test section above an acrylic wall (with mean square roughness below 0.02 µm [1], three orders smaller than our viscous length scale in our experiments). Based on experimental settings, the turbulent flow at our measurement location is considered to be a fully developed smooth-wall channel flow.

As shown in Supplementary Figure 1a, our DFRH setup consists of a 100 mW 532 nm green diode laser (OptoEngine Inc.), a 50:50 beam splitter, 10x long working distance microscopic objective (Mitutoyo Inc.), and a high-speed camera (NAC HX-5). Unlike the commonly-used digital inline holography (DIH) [2], the camera used in DFRH records the holograms generated from the interference between the backscattered light from the particles located near the wall and Fresnel reflection of light at the inner surface of the wall [3]. Note that in DFRH the outer surface of the wall is usually coated with an anti-reflection coating to minimize the reflection from this surface. Such configuration allows us to capture the signals of near-wall particles with no restriction on the tracer particle concentrations like those in conventional DIH. The detailed description of DFRH and its advantages over conventional DIH are provided in [3]. The entire DFRH setup is mounted on a vibration-damping optical table (Supplementary Figure 1b) to reduce the influence of mechanical vibration on our high-precision measurements.

All experiments are performed at a channel Reynolds number of $Re_h$=11,750 with a centerline velocity of 0.47 m/s, and friction Reynolds number $Re_\tau$=400 with a friction velocity of 0.019 m/s (calculated with the slope of the mean velocity profile within the sublayer). During our experiments, the channel is uniformly seeded with 11 μm silver coated hollow glass spheres at high concentration to ensure sufficient particles are present within the viscous sublayer for our particle tracking velocimetry (PTV) based measurements. Moreover, the particle stokes number within the viscous sublayer (calculated based on the friction velocity scale, particle size and density) is less than 0.005, sufficient to indicate high traceability of our flow with the particles. Our setup yields a spatial resolution of 1 μm/pixel over a 1×1 mm field of view with the camera operating at 3000 fps. Note that due to the smaller scattering cross section of backscattered light, the tracers appear to be around $2 \times 2$ pixels in our reconstructed holograms, different from their appearance in conventional DIH. Two types of data recording are included in our experiments, i.e., time-resolved and burst recording. For time-resolved recording, the holograms are captured continuously for 5.4 s at 3000 frames/s. In total three datasets are recorded which are used for trajectory analysis (Figure 1b-e) and acceleration analysis (Figure 3) present in the main text. For burst recording, we capture 10 frame bursts at 3000 frame/s at 50 Hz repetition (i.e., 50 bursts/s). In total nine datasets are recorded using burst mode, each of which has a sample duration of 5.4 s spread over an experimental duration of 32.3 s which are used for calculating the mean statistics (Figure 2 and Supplementary Figure 5).

**Experimental data processing**

The recorded holograms are first enhanced using a coherence based enhancement (Molaei and Sheng [4]). This approach uses cross correlation to identify holograms with matching backgrounds to calculate a time average background image for being subtracted from each hologram, suitable for the enhancement of images with backgrounds of varying intensity and noises. Subsequently, using in-house developed RIHVR algorithm [5], each hologram is reconstructed into 3D optical field of tracers and their locations are then extracted from the reconstructed field. RIHVR is an inverse reconstruction algorithm using various regularization criteria that helps limit the depth elongation of tracer particles often associated with conventional DIH reconstruction algorithms.

The extracted positions are tracked in time using both TrackPy [6,7] and an in-house developed machine learning based tracking algorithm [8] to obtain Lagrangian trajectories of particles for the following analysis. Specifically, TrackPy, based on the nearest neighbor particle matching, is first implemented to generate particle tracks. The results of TrackPy can be strongly influenced by the loss of particles in some frames due to the fluctuation in the signal strength of particles as they travel across the sample volume. Therefore, a machine learning tracking algorithm based on long short-term memory (LSTM) recurrent neural network architecture is further applied to improve the results from TrackPy. We select and manually label over 1000 tracks from the results of TrackPy and use them to train our LSTM machine learning model. The detailed information of this machine learning tracking approach is available in [8].

To obtain accurate velocity and acceleration measurements, the particle tracks are smoothed using a second order piecewise Savitzky Golay filter to eliminate any spurious high frequency fluctuations in the position data. The filter size is chosen to be 1/10$^{th}$ of each track length, which allows the window to scale with the particle velocity. The velocities are then ensemble averaged with wall normal bins of 10 μm to calculate the mean velocity profile within the viscous sublayer. The friction velocity and viscous wall unit are calculated via the slope of the mean velocity profile within $y^+$~1.2 and 3.2 and instantaneous wall shear stresses are estimated as the slope of the streamwise and spanwise velocities (i.e., $\tau_x^+ = u^+/y^+$ ; $\tau_z^+ = w^+/y^+$). To calculate the PDF of the wall shear stress, we include all data points below $y^+$~1, for both the experiment and DNS.

**Numerical simulation**

Supplementary Figure 2 shows the flow configuration used in the direct numerical simulation (DNS). The flow is periodic in the streamwise (*x*) and the spanwise (*z*) directions. No-slip and nonpermeable conditions are applied as boundary conditions in the wall-normal (*y*) direction. The simulation code uses the Fourier pseudo-spectral methods in *x* and *z* directions and Chebyshev pseudo-spectral method in *y* direction for spatial derivatives [9]. For time-advancement, we use the third-order Runge-Kutta method. All statistics denoted with "+" in this study are normalized by the combinations of the density, $\rho$, the kinematic viscosity, $\nu$, and the mean velocity gradient at the wall. The flow is driven by a constant pressure gradient *dP/dx* in the *x* direction. The flow is at the Reynolds numbers $Re_\tau$=400, where $Re_\tau$ is the friction Reynolds number. The resolutions are $\Delta x^+ = 12$, $\Delta z^+ = 7$, $\Delta y^+ = 0.5-6$. The time step size is such that the Courant–Friedrichs–Lewy (CFL) number is about 0.8. We follow [10] and average for a time duration corresponding to about $100 L_x/u_b$ to minimize sampling errors, where $L_x$ and $u_b$ are the length of the domain and streamwise velocity scale in *x* direction, respectively. We investigate the sampling errors by examining the total stress, i.e., $dU^+/dy^+ - \langle u'v' \rangle^+$ and compare it with the analytic solution (i.e., 1-*y/h*). The residual of computed total stress from the analytic solution is less than 0.006 in our DNS.

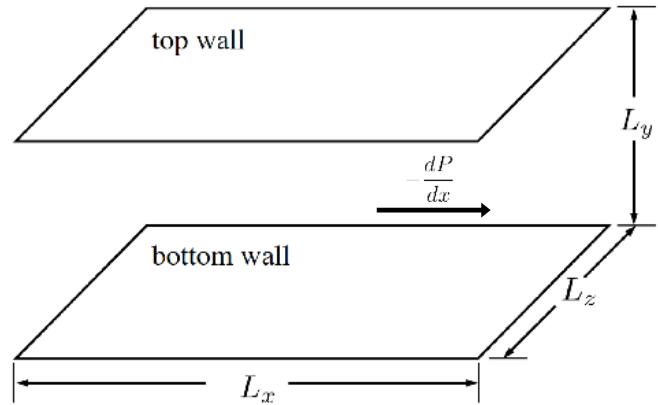

Supplementary Figure 2. A schematic showing the flow configuration used in the direct numerical simulation (DNS).

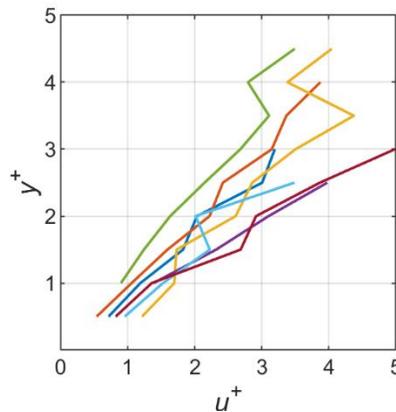

Supplementary Figure 3. Samples of instantaneous streamwise velocity profiles (different colors represent velocity profiles at different time instants) from the time resolved data showing highly unsteady non-linear distribution of velocity within the viscous sublayer.

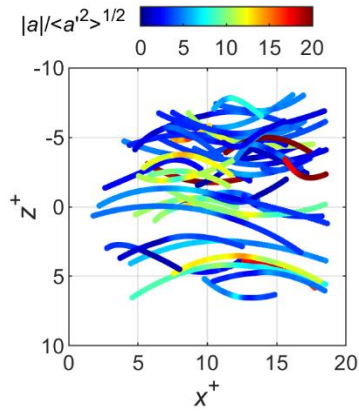

Supplementary Figure 4. Meandering trajectories colored by instantaneous acceleration magnitude illustrating strong acceleration associated with some of these tracks.

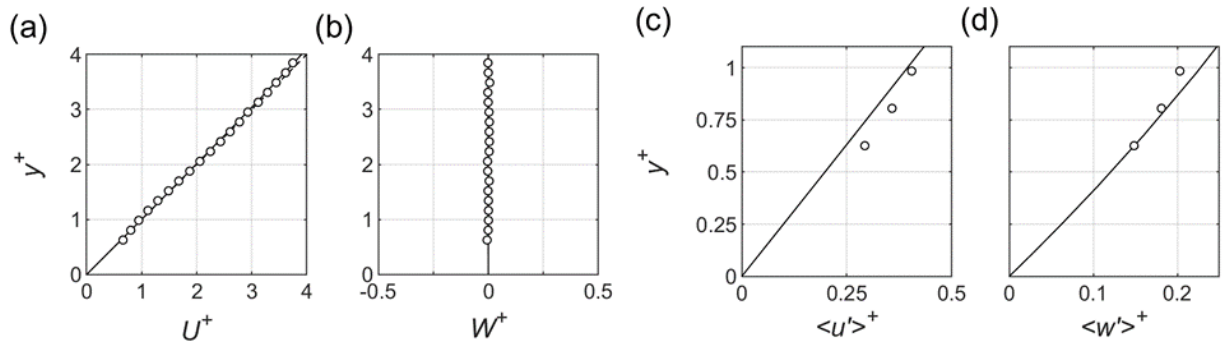

Supplementary Figure 5. The ensemble-averaged (a) streamwise and (b) spanwise velocity profiles and the (c) streamwise and (d) spanwise velocity fluctuation profiles (○ representing experimental data) compared to the corresponding DNS predictions (solid lines).

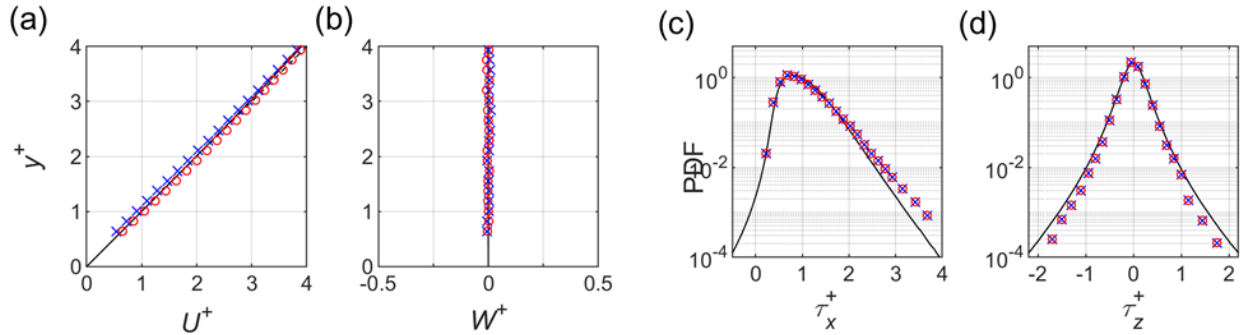

Supplementary Figure 6. Mean (a) streamwise and (b) spanwise velocity profiles and probability density functions (PDFs) of (c) streamwise and (d) spanwise wall shear stress calculated using spatially binned data based on the streamwise positions to show the independence of our measurements on streamwise positions. The quantities are calculated by splitting the data set into two groups, one with $x^+ \leq 10$ (blue ×) and the other with $x^+ > 10$ (red ○) compared to DNS values (solid lines).

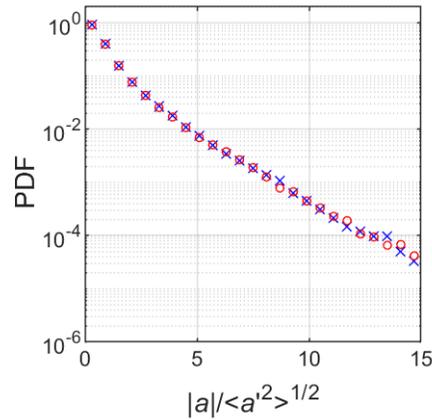

Supplementary Figure 7. Probability density function (PDF) of the magnitude of tracer acceleration calculated using spatially binned data based on the streamwise positions to show the independence of our measurements on streamwise positions. The quantities are calculated by splitting the data set into two groups, one with $x^+ \leq 10$ (blue ×) and the other with $x^+ > 10$ (red ○).